\documentclass[journal]{IEEEtran}

\usepackage{amsfonts}
\usepackage{amsmath}
\usepackage{amssymb}
\usepackage{url}
\usepackage{graphics}
\usepackage{graphicx}
\usepackage{enumitem}

\newcommand{\be}{\begin{equation}}
\newcommand{\bea}{\begin{eqnarray}}
\newcommand{\ee}{\end{equation}}
\newcommand{\eea}{\end{eqnarray}}
\newcommand{\nn}{\nonumber}

\newcommand{\qg}{\gamma}

\newcommand{\qe}{\varepsilon}

\newcommand{\qf}{\varphi}

\newcommand{\pr}{{\rm Pr}}
\newcommand{\fr}[2]{{\textstyle \frac{#1}{#2}}}

\newcommand{\EE}{{\mathbb E}}

\newcommand{\bits}{ \{0,1\} }
\newcommand{\cA}{{\mathcal A}}

\newcommand{\cC}{{\mathcal C}}
\newcommand{\cD}{{\mathcal D}}
\newcommand{\cE}{{\mathcal E}}

\newcommand{\cO}{{\mathcal O}}

\newcommand{\cS}{{\mathcal S}}
\newcommand{\cT}{{\mathcal T}}

\newcommand{\cV}{{\mathcal V}}

\newcommand{\stirling}[2]{ \left\{ {#1 \atop #2} \right\} }
\newcommand{\isdef}{\stackrel{\rm def}{=}}

\begin{document}

\title{The Blob: provable incompressibility and traceability in the whitebox model}

\author{
Boris \v{S}kori\'{c}, Wil Michiels
}

\markboth{ }{header2}

\maketitle

\begin{abstract}
We introduce a scheme for distributing and storing software with cryptographic functionality 
in the whitebox attacker model.
Our scheme satisfies two relevant properties: incompressibility and traceability.
The main idea is to store a large amount of random data (a `blob'), some of which will be randomly sampled
in the future to serve as key material, and some of which serves as a watermark.
We study two variants: with and without re-use of key material.
For both variants we analyse how many decryptions can be performed with the blob,
taking into account collusion attacks against the watermark. 
Our results show that application of blob schemes in the context of pay-TV is feasible.
\end{abstract}

\newtheorem{theorem}{Theorem}
\newtheorem{lemma}{Lemma}


\section{Introduction}
\label{sec:intro}

\subsection{Traceable big keys}

The use of extremely large cryptographic keys (``Big-Key Cryptography'', BKC) 
\cite{DLL2005,CLW2006,CDDLLW2007,BKR2016}
has been studied as a countermeasure against key exfiltration from infected devices.
If the adversary's bandwidth is limited, an assumption known as the Bounded Retrieval Model,
BKC ensures that
the exfiltrated material will be smaller than a whole key and hence insufficient to 
perform cryptographic operations.
Another security measure that consumes large amounts of memory is {\em White-Box Cryptography}
(WBC or `whitebox') \cite{CHES2017CTF,CEJvODRM2002,CEJvOSAC2002}, the art of obfuscating a symmetric encryption/decryption executable such 
that the key is difficult to extract.
The difference between these two techniques is that a WBC executable contains a short key 
(e.g.~a single AES key) in a blown-up form, whereas in BKC the key itself is very large.

In this paper we consider BKC in combination with {\em traceability}.
A group of users receive software that contains exactly the same decryption functionality
but which is differently watermarked for each user.
If a user leaks his software, the origin of the leak can be traced. 
As example use cases for this kind of group functionality with software traceability one
can think of
(a) protection of pay-TV keys, many of which are shared by multiple customers; and
(b) anonymous credentials, where only group membership is verified without revealing a person's identity.

These scenarios of course do not require Big-Key Cryptography.
In pay-TV WBC has been deployed \cite{CEJvODRM2002}, which is also well suited to watermark-based tracing.
Anonymous credentials systems have been constructed using {\em group signatures} \cite{BMW2003,BBS2004,CDLNT2020};
tracing of leaked credentials is easy since the users' keys are different.
BKC however, when used properly, has interesting advantages over these solutions.
For instance, in typical group signatures the crypto and the tracing are not post-quantum secure.
In contrast, we will present a BKC tracing scheme that is {\em information-theoretically secure},
and it is easy to make the decryption post-quantum secure.
Furthermore, in WBC one of the principal properties looked after 
is incompressibility: it should be infeasible to compress the obfuscated block cipher executable.
Most whitebox techniques are either completely broken or are known to be vulnerable to attacks
inspired by side channel analysis \cite{AMR2019,ABBHMMSTT2019}. 
On top of that, whitebox traceability is no stronger than its incompressibility.
In contrast, the incompressibility of BKC is information-theoretic.

In this paper we present and analyse a BKC scheme with traceability based on a binary Tardos code.
We focus on a pay-TV use case.

\subsection{Related work}
\label{sec:related}

\underline{Big-Key Cryptography}.\\
Over the last two decades various schemes have been developed
to protect against exfiltration of data in general \cite{DLL2005}
and keys in particular \cite{CLW2006,CDDLLW2007,BKR2016}.
The Bounded Retrieval Model (BRM) assumes that the total amount of data that the attacker can exfiltrate is limited.
Under the BRM assumption security is provided by working with keys and data structures whose size exceeds this limit.
To avoid paying a speed penalty on top of the memory cost,
cryptographic operations are then based on (pseudo)randomly chosen subkeys instead of the whole key.
Using extractors it is then possible to derive keys that are uniform even to an attacker who has
observed parts of the big key \cite{BKR2016}.
We will not go to this level of sophistication but simply count how many key bits are known 
to the adversary. 
An important difference between our paper and \cite{BKR2016} is that
we don't work in the BRM.
In our case the adversary's strategy is dictated not by the wish to extricate exactly the right data
but by the necessity to stay untraceable.

\underline{Broadcast encryption}.\\
The aim of {\em broadcast encryption} is to send different data to different recipients
over a broadcast channel.
One example is customers of a pay-TV operator who have different subscriptions,
i.e.\,get access to a different subset of TV channels.
A possible solution is to encrypt the transmissions according to a so-called {\em revocation tree} \cite{FN1993,NNL2001}.
The customers are the leaves of the tree. 
Each node of the tree has a node key associated with it.
Each customer has a decoder device which comprises a number of node keys, 
for instance those lying on the path from the leaf to the root.
By appropriately encrypting a message with a specific subset of the node keys and broadcasting the ciphertexts,
the operator can specify in a fine-grained way which leaves are able to recover the plaintext.
In this way the operator can handle not only multiple subscription types but also
revocation of customers who have misbehaved.
Revocation trees are well suited for `static' use cases, i.e.~situations where the device keys cannot easily be replaced.
In more dynamic environments broadcast encryption can be achieved without revocation trees.

In the pay-TV setting different forms of unauthorized redistribution (`piracy') exist.
One method is for the attackers to publish node keys from decoder devices.
As long as these keys are from nodes close to the root, they are present in many decoders and hence 
they do not help the operator to pinpoint who the pirates are.
After this kind of key compromise, the operator can recover by using node keys closer to the leaves;
the compromised keys may even be refreshed in such a way that only authorized users receive the replacement keys.
Depending on the position of the leaked keys in the tree, the recovery may take a lot of bandwidth.

A second form of piracy is the re-broadcasting of decrypted content.
Here the countermeasure is {\em content watermarking}.
The operator puts different watermarks into the content streams, which then get encrypted with different node keys;
in this way it becomes visible from the pirate stream which decryption key was employed by the attackers.
For the operator this kind of tracing is expensive in terms of bandwidth, since multiple instances of the same content 
have to be broadcasted simultaneously.
The practical solution is to send a single stream most of the time,
and only occasionally duplicate a small piece of the stream; 
a `0' watermark is sent to one subset of the customers and a `1' watermark to the rest.
By varying the composition of the customer subset for each duplicated piece,
the operator can zoom in on individual leaf keys.
When only one compromised decoder box is used in the re-broadcasting attack,
the watermarking technique allows the operator to identify the pirate relatively quickly.
When the keys from $c$ boxes are combined (a {\em collusion}), the required number of steps scales as $c^2$ \cite{Tardos}.
Special codes have been developed that resist collusion attacks,
in particular a class of bias-based codes commonly known as Tardos codes \cite{Tardos}.
For an overview we refer to~\cite{Skoric2016}.
Tardos codes have been proposed to watermark large data structures in the context of combined client-side watermarking 
and decryption (`fingercasting')~\cite{KSCS2007}.

In this paper we will focus on a broadcast encryption scenario that requires a revocation tree,
and we propose to use a Tardos code to watermark a BKC implementation of the top-level node keys.

\underline{White-Box Cryptography}.\\
The term `white-box attacks' refers to attacks where the adversary observes all details of a program's execution.
The aim of WBC is to create executables with cryptographic functionality
in such a way that several security properties are satisfied even in the white-box attacker model.
Obfuscation techniques typically result in very large executables,
which are called White-Box Implementations (WBIs) \cite{CEJvODRM2002,CEJvOSAC2002}.
One of the desirable security properties is {\em incompressibility}:
even if it is not possible to prevent the attacker from running a copy of the executable on another device,
it should at least
not be too easy for the adversary to create a version of the attacked WBI that is significantly smaller 
than the WBI.\footnote{
An extreme case is e.g.~a WBI that implements AES-encryption / decryption; if the attacker learns the key, the WBI can be compressed to
128 bits.
}

A second desirable property is {\em traceability} \cite{Michiels2010}.
Consider a key that multiple parties possess, e.g.\,a node key as described above.
If the individual WBIs of the node key are differently {\em watermarked}, then a leaked node key can be traced.
However, if the WBI is compressible then the watermark can be destroyed with high probability.

WBIs are often made under the constraint that the cryptographic functionality of the WBI must equal that of
a standardised block cipher, e.g. DES or AES. 
Such a constraint usually comes from regulations or compatibility requirements.
Published white-boxing techniques that obfuscate a block cipher do not provide
provable incompressibility.

\vskip2mm

\underline{Anonymous credentials}.\\
In a group {\em signature} scheme, the enrolled parties receive credentials (signing keys)
which differ from each other but which allow each party to sign data in such a way that
the signature does not reveal which member of the group created the signature.
Schemes with short signatures have been obtained \cite{BBS2004,CDLNT2020}
using bilinear maps.
Typically there is a mechanism by which the anonymity can be revoked,
either by the group `manager' or collectively by the members \cite{CSST2006}
via some threshold mechanism.
This revocation occurs when a group member has misbehaved in some way,
and is based on the signature(s) without access to the credential itself. 
For the purpose of the current paper the privacy revocation mechanism is not relevant;
instead we care about {\em the credential itself} being 
(i) exfiltrated by an attacker or 
(ii) widely shared by the group member.  
Furthermore, in the context of group signatures there is the concept of collusion resistance,
meaning that a collusion of group members should not be able to create a signature that implicates another member.
This is a different concept from the collusion-resistant watermarking that we will use.

Finally, we use the term `anonymous credential' for anything that allows a user to prove group membership,
regardless of the technique; 
in our BKC scheme the user demonstrates his ability to decrypt a (symmetric-crypto) ciphertext.

\vskip2mm

\underline{Comparison to related work}.\\
Summarised very briefly, this paper differs from related work as follows.

\vskip1mm

\begin{tabular}{|l||l|}
\hline
{\it Compared to ...} & {\it our scheme has ...} \\
\hline
Big Key Crypto & tracing \\
\hline
whitebox & provable incompressibility and tracing; 
\\
 & no speed penalty on decryption \\
\hline
revocation trees & tracing of high-level node keys; 
\\ &
defense against exfiltration \\
\hline
group signatures & post-quantum security of the crypto; 
\\ &
unconditionally secure tracing; \\
& defense against exfiltration \\
\hline
\end{tabular}

\subsection{Contributions and outline}
\label{sec:contrib}

We propose a very simple BKC scheme that has incompressibility and traceability
in the white-box attacker model.
We construct a decryption algorithm based on
a large lookup table (a {\em blob}) whose entries are randomly probed to provide key material.
Interspersed with the functional entries there are entries that contain tracing data.
The blobs belonging to different users contain exactly the same functional entries but different tracing entries.

The blob is incompressible because all of its functional entries may be used in the future, 
and the attacker cannot predict the access sequence.
The blob is traceable because the attacker cannot sufficiently distinguish between functional entries and
tracing entries.
Furthermore, a Tardos code is applied to ensure collusion resistance.

We have two versions of our scheme: 
(i) single use of blob entries, i.e.\,an entry gets discarded after it has been used;
(ii) multiple use.
Both versions are easy to analyse with basic combinatorics.
An important design parameter is the width (in bits) of each blob entry;
this influences the number of entries that need to be addressed in order to build one decryption key.
As a figure of merit we use the number of times `$n_{\rm max}$'
that a blob can probed
before the colluders have enough information to create an untraceable pirated blob.

\begin{itemize}[leftmargin=4mm,itemsep=0mm]
\item
In Section~\ref{sec:withoutreuse} we give a step-by-step description of the {\em single-use} variant,
and we analyse its properties.
It turns out that it is optimal to split each key into individual bits, which are picked from different locations in the blob.
Furthermore, we find that the figure of merit $n_{\rm max}$ depends on the number of colluders only weakly.
\item
In Section~\ref{sec:withreuse} we introduce and analyse the {\em multiple-use} scheme variant .
Here the optimal width of blob entries equals the key size, i.e.\;keys are not split up.
We find that the $n_{\rm max}$ quickly degrades with increasing number of colluders. 
\item
We observe a cross-over as a function of the coalition size:
When the number of colluders is small, the multiple-use scheme performs best,
whereas the single-use scheme is best in the case of large coalitions.
This is illustrated in Fig.\ref{fig:sketch} below.
\end{itemize}

\begin{figure}[h]
\begin{center}
\includegraphics[width=80mm]{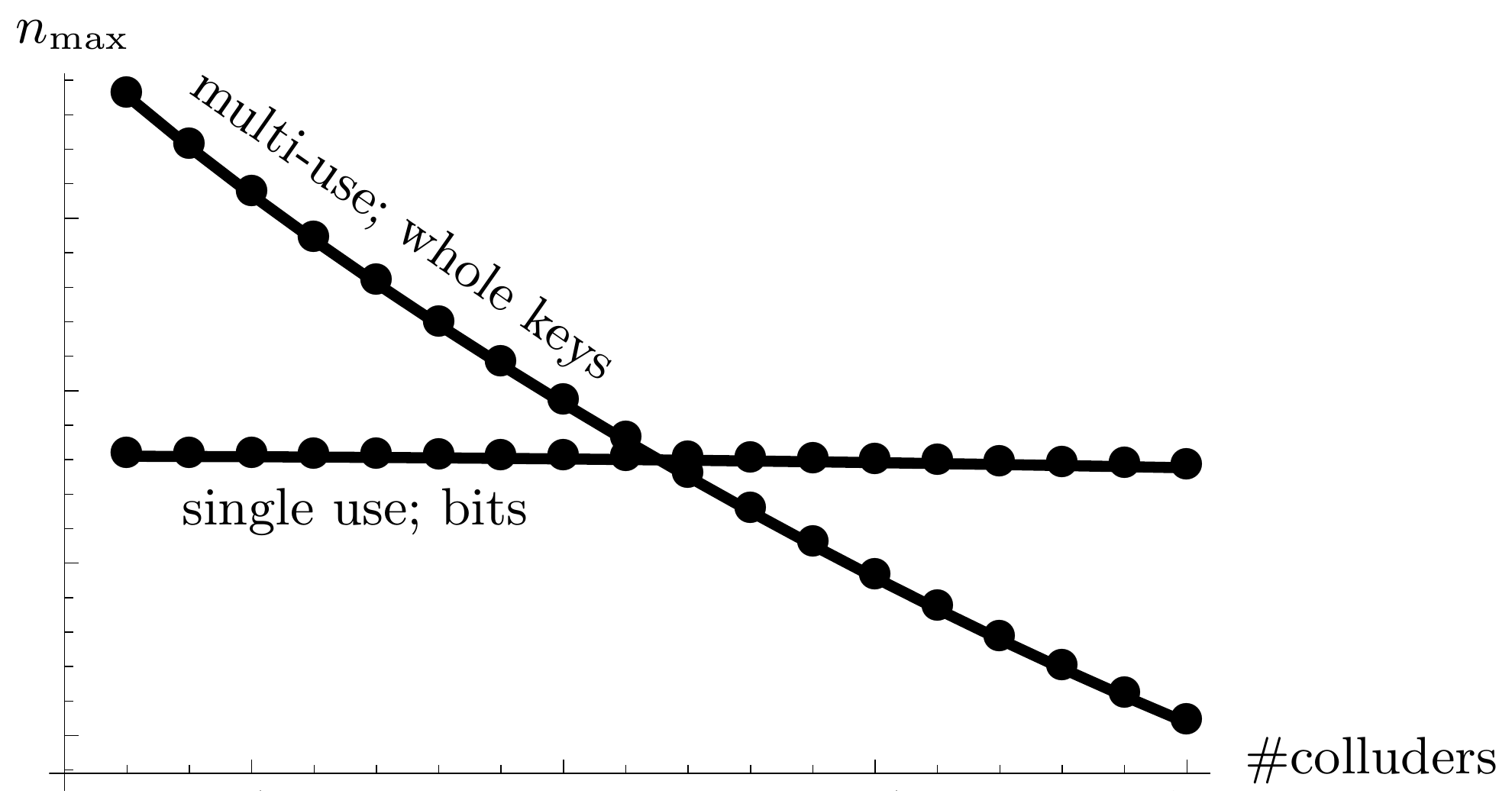}
\end{center}
\caption{\it
Artist impression of the crossover between the two types of blob scheme.
(See Fig.\ref{fig:combo} for actual numbers.)
The graph plots the maximum number of blob uses as a function of the number of colluders.
Shown is the single-use scheme with blob entries that are single bits,
and the multi-use scheme with blob entries that are full keys (128 bits). 
}
\label{fig:sketch}
\end{figure}

In the context of content distribution, our scheme may be applied to watermark a number of high-level node keys
in the revocation tree.
This has the advantage that node key publishing attacks, which are especially powerful for those high-level nodes, get thwarted.
In Section~\ref{sec:payTV} we show that the typical parameters in pay-TV (e.g. number of users, key refresh rate) are consistent with the use of a blob scheme.

Compared to Whitebox watermarking, the advantage of a blob scheme is information-theoretic security of the tracing.
Furthermore a blob scheme has provable (information-theoretic) incompressibility.

In the context of anonymous credentials, a blob scheme can be used to mimic the effect of group signatures.
The ability to perform a {\em symmetric} decryption replaces the ability to create an {\em asymmetric} group signature.
The advantages of a blob are (i) post-quantum security of the crypto; (ii) unconditional security of the tracing.
A disadvantage is of course the size of the software.

\section{Preliminaries}

\subsection{Notation and terminology}
\label{sec:notation}

We use the notation $[N]$ for $\{1,\ldots,N\}$.
We write sets in calligraphic font.
For a vector $v$ and a set $\cA$, the vector $v_\cA$ is defined as $(v_j)_{j\in\cA}$.
For vectors $v$ and $w$, the vector $v_w$ is defined as $(v_{w_1},v_{w_2},\ldots)$.
An erasure symbol (empty output) is written as $\bot$.
The falling factorial is denoted as $(x)_k=\frac{x!}{(x-k)!}$.
Stirling numbers of the second kind are denoted as 
$\stirling{n}{k}$, with the property $x^n=\sum_{k=0}^n \stirling{n}{k}(x)_k$.
The number $\stirling{n}{k}$ counts how many ways there are to partition the integer $n$ into 
$k$ (unlabeled) nonzero parts.

We will consider a broadcasting context with an Operator who manages all the keys and
prepares all encryptions, and which is also the party that performs the tracing.
The number of users is $U$. The users are labeled $1,\ldots,U$.
The coalition (set of colluders) is $\cC\subset[U]$, with $|\cC|=c$.

In broadcast encryption the content is encrypted with a {\em content key} that is occasionally refreshed.
The broadcast contains special parts called {\em control messages} that allow each authorized user 
to obtain the content key.

We define the function BinoTail as the probability mass in the right tail of 
a binomial distribution,
${\rm BinoTail}(\ell,a,p)\isdef\sum_{j=a}^\ell{\ell\choose j} p^j (1-p)^{\ell-j}$.
The inverse function with respect to the third argument is denoted as InvBinoTail.

\subsection{Tardos codes}
\label{sec:prelimTardos}

The study of collusion-resistant codes has resulted in 
a good understanding of fingerprinting capacities
and in efficient codes that approach capacity.
For the purposes of this paper it suffices to quote one result from the literature,
namely the sufficient code length for tracing at least one member of the attacking
coalition.

The watermark is abstractly modelled as a a string of watermark symbols from some alphabet,
embedded in various `positions' in the content.
An {\em undetectable position} is defined as a position where all colluders receive the same symbol.
Collusion attacks are often modelled using the Marking Assumption (MA) \cite{BonehShaw}:
in undetectable positions the colluders are allowed to output only the symbol that they have observed.
Let $P_{\rm FP}$ be the overall probability that a false accusation occurs in the tracing procedure.
When binary codes are used, the sufficient code length as a function of $c,U,P_{\rm FP}$ is given by
\be
	L^{\rm binary}_{\rm MA,suff}(c,U,P_{\rm FP}) \approx \frac{\pi^2}2 c^2 \ln\frac{U}{P_{\rm FP}}.
\label{codelength}
\ee
In this paper we restrict ourselves to binary codes because a larger alphabet would 
open up powerful attack avenues and/or make the blob scheme less efficient.\footnote{
In detectable positions, the colluders have the option of creating random data, which (depending on the width of blob entries)
may likely collide with existing watermark symbols. Such collisions are detrimental to tracing.
Furthermore, for non-binary watermarking alphabet it becomes impossible to reduce the blob entry width to one bit.
}

If the Marking Assumption is relaxed a bit, by allowing erasures $\bot$ in a fraction $\qe$
of all undetectable positions, then the code length (\ref{codelength}) increases by a factor $(1-\qe)^{-2}$ \cite{SKSC2011}.

\section{Blob scheme without re-use of entries}
\label{sec:withoutreuse}

\subsection{Scheme description}

The general idea of our scheme is that the decryption-enabling data structure consists of a large (pseudo)random data `blob',
with tracing information inserted at positions that are known only to the Operator.
Let $N$ be the number of data entries in the blob~$B$.
We write $B=(B_i)_{i=1}^N$, with $B_i\in\bits^w$.
The size of the blob (in bits) is denoted as $M$,
\be
	M = Nw.
\ee
A $k$-bit decryption key is obtained by collecting data from $\ell$ positions in the blob,
\be
	k = \ell w.
\ee
In one extreme case $w=1$ and $\ell=k$, i.e. an entry is a single bit.
In another extreme case $w=k$ and $\ell=1$, i.e. each blob entry is an entire key. 
The set of tracing positions is $\cT\subset[N]$, with $|\cT|=t$.
In the positions $\cT$ the Operator implements a binary Tardos code;
for $j\in\cT$ only two different values $B^{(0)}_j,B^{(1)}_j\in\bits^w$ can be handed out to the users.

Below we outline the essence of the scheme.
The approach works for any symmetric\footnote{
Or asymmetric. But then the advantage of post-quantum security may be lost.
}
decryption algorithm, and therefore we do not specify the cipher.

\underline{Initialisation}\\
The Operator (pseudo)randomly generates $\cT$ and $B_{[N]\setminus\cT}$. This data is the same for all users.
For each user $u$ individually he inserts tracing information $z^{(u)}$ in $(B_j)_{j\in\cT}$.
The personalised blob $B^{(u)}$ is given to user~$u$. 
The operator remembers $z^{(u)}$ for all $u\in[U]$, and $\cT$ and $B_{[N]\setminus\cT}$.
The Operator initializes a set $\cV$ to the empty set, $\cV=\emptyset$.
The purpose of this set is to keep track which blob entries have been used up.

\underline{Encryption}\\
The Operator (pseudo)randomly generates a vector $L=(L_1,\ldots,L_\ell)\in([N]\setminus(\cT\cup\cV))^\ell$.
It must hold that $L_i\neq L_j$ for $i\neq j$.
He encrypts the content using $B_L$ as key material.
He broadcasts the ciphertext and a compact description of~$L$ to all users.
He updates $\cV\mapsto\cV\cup\{L_1,\cdots,L_\ell\}$.

\underline{Decryption}\\
All users receive the ciphertext and the description of~$L$.
User $u$ decrypts the ciphertext using key material $B^{(u)}_L$.

\vskip3mm

The encyption and decryption phase are repeated many times before a new initialisation is required.
Note that the users do not have to store~$\cV$; only the operator does.
The incompressibility is a consequence of the following facts,
\begin{itemize}[leftmargin=4mm,itemsep=0mm]
\item
It takes many iterations before a user can guess $\cT$ with any accuracy.
\item
If user $u$ publishes an edited blob $\tilde B^{(u)}$ that contains a substantial part of $B^{(u)}_\cT$,
he is traceable.
On the other hand, if he damages the non-tracing part of the blob, the published $\tilde B^{(u)}$ will lose
some of its decryption-enabling functionality.
\end{itemize}

\underline{Control messages}\\
Broadcasting a description of $L$ is potentially expensive in terms of bandwidth.
In the worst case, the Operator draws $\ell$ truly random pointers $L_1,\cdots,L_\ell$
under the constraint $L_i\notin \cT\cup\cV$;
then he has to broadcast $L\log N$ bits.
In the most optimistic case, the Operator generates $L_1,\cdots,L_\ell$ pseudorandomly from a single seed;
then only the seed has to be broadcast.
Note that in this approach the constraint $\forall_{i\in[\ell]}\; L_i\notin\cT\cup\cV$ is not automatically satisfied.
Multiple seed values have to be tried before the constraint is satisfied.
Depending on $\ell$ the trial-and-error procedure may be too cumbersome for the Operator, and it may be 
preferable to derive $L$ from more than one seed.

\vskip1mm

{\small
\begin{tabular}{|l|l|}
\hline
$M$ &  Size of the blob, in bits.
\\ \hline
$w$ & Word size of the blob entries, in bits.
\\ \hline
$N$ & Number of blob entries. $M=Nw$.
\\ \hline
$B$ & The Blob. $B=(B_i)_{i=1}^N$, $B_i\in\bits^w$.
\\ \hline
$t$ & Number of blob entries used for tracing.
\\ \hline
$\cT$ & The set of indices where $B$ contains tracing information. 
\\ \hline
$k$ & Key size in bits.
\\ \hline
$\ell$ & Number of blob entries needed to build a key. $k=\ell w$.
\\ \hline
$\cV$ & Set of used blob indices.
\\ \hline
$n$ & Number of times the blob has been used.
\\ \hline
$\qe$ & Fraction of data thrown away by the attackers.
\\ \hline
$\qe_*$ & Threshold value for $\qe$. 
\\ \hline
$U$ & Number of users.
\\ \hline
$\cC$ & Coalition of users. $\cC\subset[U]$.
\\ \hline
$c$ & Size of the coalition.
\\ \hline
$c_0$ & Size of the coalition anticipated by the Operator.
\\ \hline
$P_{\rm FP}$ & False Positive probability (false accusation).
\\ \hline
$L_{\rm suff}$ & Sufficient code length for a binary Tardos code.
\\ \hline
$k_0$ & Key size considered `sufficiently difficult' to brute-force.
\\ \hline
$\qg$ & Tolerated prob. that pirated blob fails to provide next key. 
\\ \hline
\end{tabular}
}

\subsection{Attack description}
\label{sec:attacksingleuse}

The coalition is a set $\cC\subset[U]$ of users, with $|\cC|=c$.
The output of the attack is denoted as $y\in(\bits^w\cup\bot)^N$.
The colluders compare their versions of the blob, $(B^{(u)})_{u\in\cC}$.
In some positions $\cD\subset[N]$ they notice a difference. These positions are called `detected' positions.
The set of detected symbols in position $i$
is denoted as $\cS_i\subseteq \bits^w$.
In any position the colluders are allowed to output any value in $\bits^w$ or an erasure.\footnote{
This is very different from audio-video watermarking, where the attack must not cause 
perceivable glitches.
}
For $i\in\cT$, outputting $y_i\notin \cS_i$ carries some risk since it can happen
that $y_i$ is not in the (position-dependent) binary alphabet;
such a `symbol error' informs the Operator that the colluders have different data in position~$i$,
which helps him a lot.
The probability of causing a Symbol Error is
$\frac{2^w-2}{2^w-|\cS_i|}$.
Because of this risk to the colluders we will not consider random outputs.

We consider an attack that consists of two steps,
\begin{enumerate}[leftmargin=5mm,itemsep=0mm]
\item
An `ordinary' collusion attack on the unused part of the blob, $[N]\setminus\cV$, 
under the Marking Assumption,
resulting in a new blob~$\tilde B$
which may differ from the original ones only in the detected positions~$\cD$.
\item
The colluders select a random subset $\cE\in([N]\setminus\cV)\setminus\cD$
of size $|\cE|=\qe(N-|\cV|-|\cD|)$.
The output of the attack equals $\tilde B$ with the replacement $\tilde B_j=\bot$ for all $j\in\cE$.
\end{enumerate}

The purpose of the second step is to reduce the traceability of the colluders
in case the first step is not enough.
(Note that we assume the system parameters $t,L_{\rm suff}$ to be publicly known
in accordance with the Kerckhoffs principle.
The colluders know exactly when they have achieved untraceability.)
An unintended side effect, from the point of view of the attackers,
is that they are destroying key material contained in $B_\cE$.

\subsection{Analysis of the attack}
\label{analysissingle}

Let $n$ be the number of keys that have been used up.
Then $|\cV|=n\ell$. 
We introduce the notation $k_0$ for the lowest key length that is considered
`difficult' to brute-force.
Furthermore we introduce a parameter $\qg$ which represents a target probability for
the attack to fail.
We consider $k_0,\qg$ to be public, in accordance with the Kerckhoffs principle.

As the figure of merit for a scheme at fixed $M,k,k_0,\ell,\qg,L_{\rm suff}$
we will use $n_{\rm max}$.
The $n_{\rm max}$ represents the number of blob uses before it becomes possible
for the attackers to publish an untraceable pirate blob which allows for `easy'
brute-forcing of future keys (fewer than $k_0$ bits) with probability $\geq 1-\qg$.

\subsubsection{Traceability}
\label{sec:tracingsingleuse}

The deletions in $\cE$ have the effect of erasing a fraction
$\approx\qe$ of the {\em undetectable} positions in~$\cT$.
This can be seen as a modification of the attacker model to incorporate noise.
As mentioned in Section~\ref{sec:prelimTardos},
for this modified attacker model the 
required code length is increased by a factor $(1-\qe)^{-2}$.  
If $(1-\qe)^{-2}> t/L_{\rm suff}$ then the Operator no longer has 
control over the False Accusations; effectively the colluders are untraceable.
The breakeven point lies at
\be
	\qe_*= 1- \sqrt{\frac{L_{\rm suff}}{t}}.
\ee

If the traitor tracing scheme is chosen badly, then $L_{\rm suff}$ is large;
then there is not much room in the blob for $t$ to be much larger than $L_{\rm suff}$, resulting in a small value
of $\qe_*$.

\begin{figure}[h]
\begin{center}
\includegraphics[width=65mm]{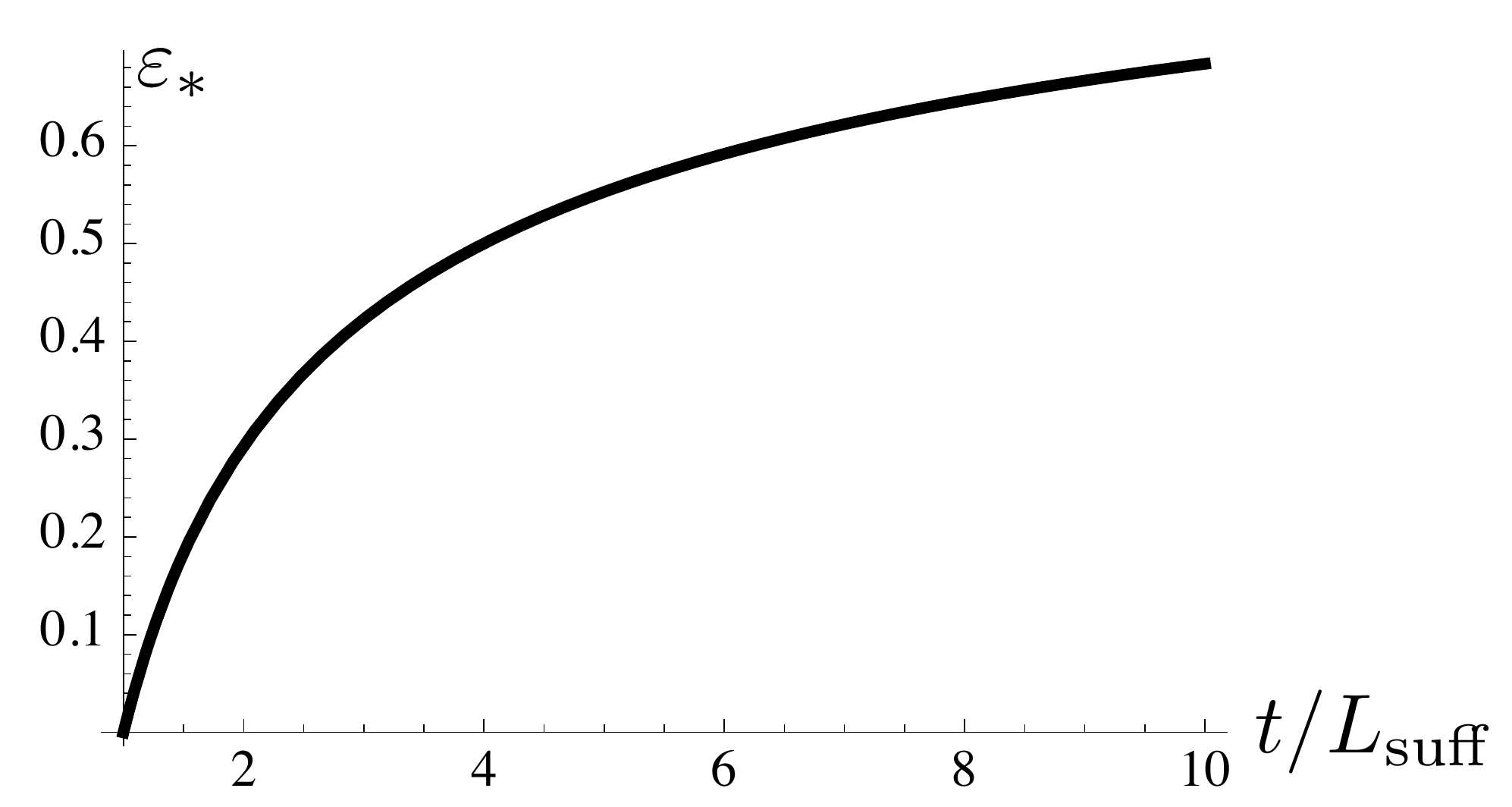}
\end{center}
\caption{ {\it
Threshold value $\qe_*$ of the thrown-away fraction $\qe$
as a function of the number of tracing positions.
In order for the pirates to remain undetected they have to set $\qe>\qe_*$.
}}
\end{figure}

\subsubsection{Probability that the pirated blob fails to provide the next key}

Brute-forcing fails if at least $\lceil k_0/w\rceil$ blob entries
out of $\ell$ are not available to the attackers.
When a fraction $\qe_*$ of the functional entries is missing,
the probability of this event is computed as a partial binomial sum as follows,
\bea
	\pr[\mbox{\#unavailable bits}\geq k_0] 
	&=& \sum_{a=\lceil k_0/w\rceil}^\ell {\ell\choose a}\qe_*^a(1-\qe_*)^{\ell-a}
	\nn\\
	&=& {\rm BinoTail}(\ell,\lceil\fr{k_0}{w}\rceil,\qe_*).
\label{prob_singleuse}
\eea
A special case occurs for $w\geq k_0$: the attackers need all $\ell$ chunks and 
(\ref{prob_singleuse}) reduces to 
$
	\pr[\mbox{\#unavailable bits}\geq k_0] = 1-(1-\qe_*)^\ell.
$

The Operator is happy if the probability (\ref{prob_singleuse}) exceeds the design parameter~$\qg$.
At given $k,k_0,\ell,\qg$
the borderline case is setting $t$ such that 
\be 
	{\rm BinoTail}(\ell,\lceil\fr{k_0}{k}\ell\rceil,\qe_*)=\qg.
\label{t_noreuse}
\ee
The $t$ enters via $\qe_*=1-\sqrt{L_{\rm suff}/t}$.
\be
	t = L_{\rm suff}\Big[1-{\rm InvBinoTail}(\ell,\lceil\fr{k_0}{k}\ell\rceil,\qg)\Big]^{-2}
\ee
Note that (\ref{t_noreuse}) does not depend on $M$ and~$n$.

\subsubsection{Figure of merit}

The figure of merit $n_{\rm max}$ is easy to obtain as a function of $M,k,k_0,\ell,\qg,L_{\rm suff}$,
since the attackers' success probability does not actually depend on~$n$.
The $n_{\rm max}$ simply equals the total number of keys stored in the blob.
However, this number depends on $k,k_0,\ell,\qg,L_{\rm suff}$ nontrivially via the parameter~$t$.
On the one hand, the amount of key material contained in the blob (in bits) is $M-tw$.
On the other hand, each blob use consumes $k$ bits. The number of uses required to use up all
key material is given by
\bea
	&& \hskip-4mm
	n_{\rm max}^{\rm single}(M,k,k_0,\ell,\qg,L_{\rm suff})
	=
	\frac{M-tw}{k} = \frac Mk-\frac t\ell
	\nn\\&& \quad = 
	\frac Mk-\frac 1\ell
	L_{\rm suff}\Big[1-{\rm InvBinoTail}(\ell,\lceil\fr{k_0}{k}\ell\rceil,\qg)\Big]^{-2}.
	\quad
\label{nmaxsingleuse}
\eea

\subsubsection{Choosing the parameters}

\begin{figure}[h]
\begin{center}
\includegraphics[width=70mm]{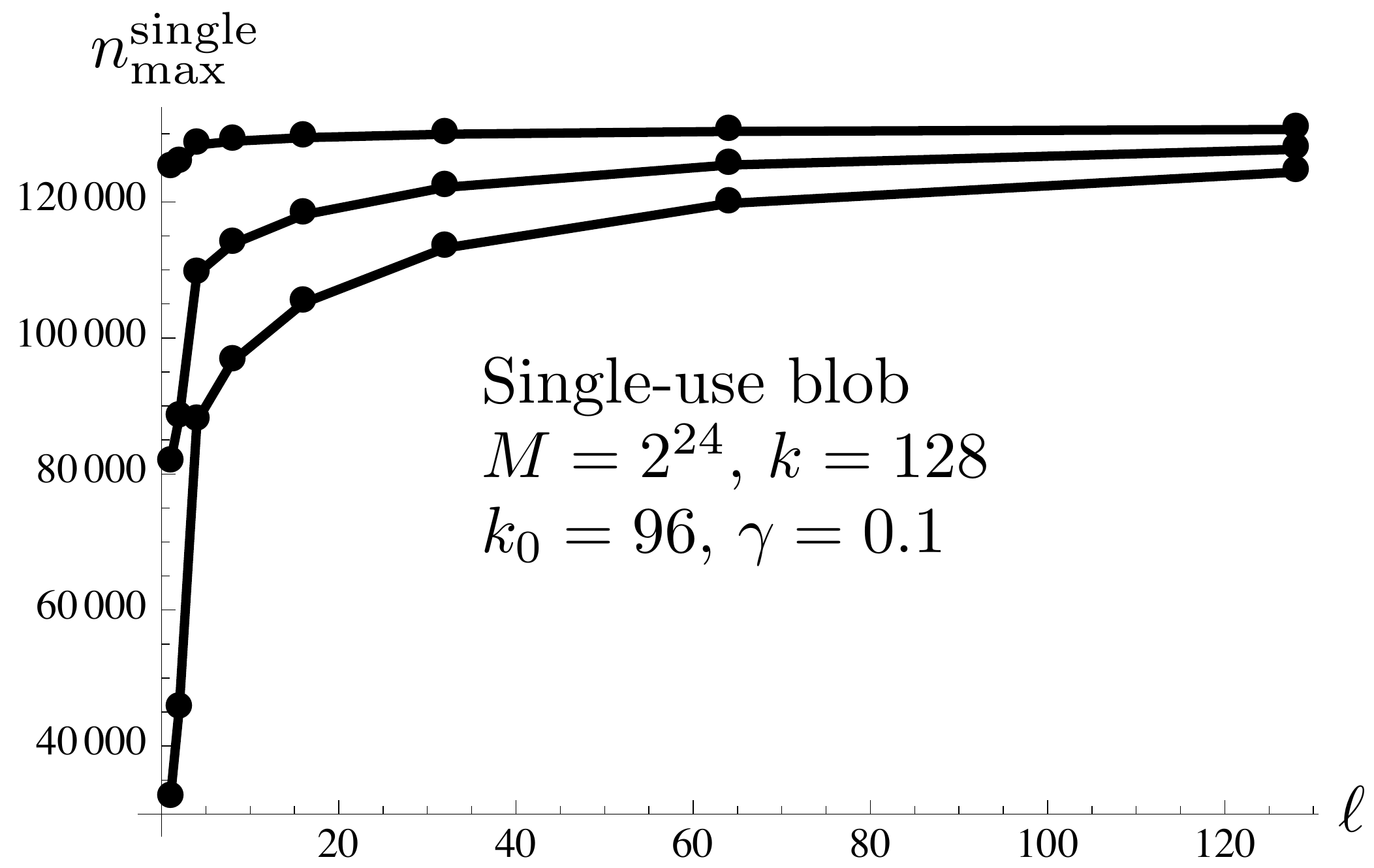}
\end{center}
\caption{ {\it
$n_{\rm max}^{\rm single}$ (\ref{nmaxsingleuse}) as a function of $\ell$, at
$M=2^{24}$, $k=128$, $k_0=96$, $\qg=0.1$. 
From top to bottom: $L_{\rm suff}=5000,40000,80000$. 
}}
\label{fig:nmaxsingleell}
\end{figure}

\begin{figure}[h]
\begin{center}
\includegraphics[width=70mm]{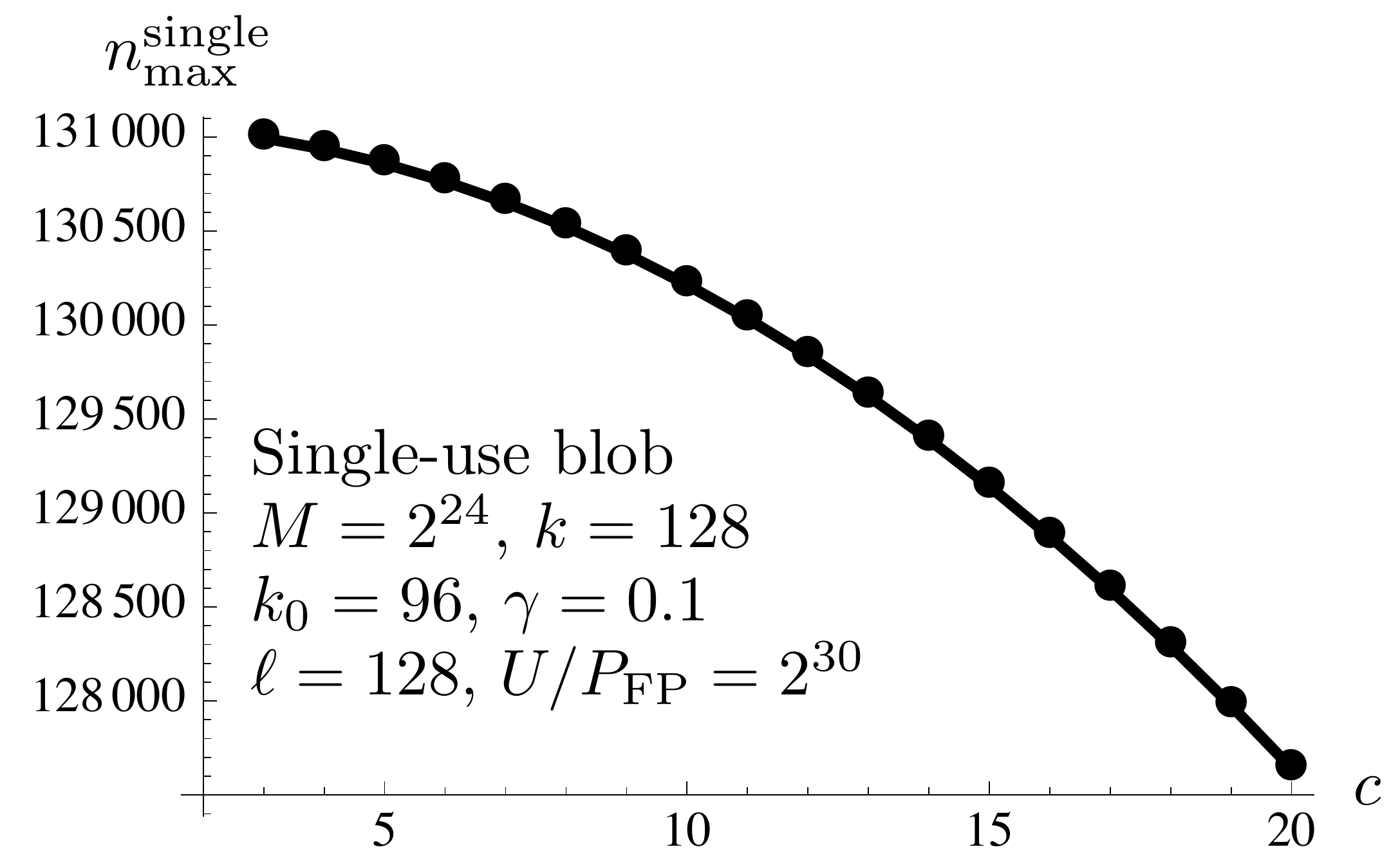}
\end{center}
\caption{ {\it
$n_{\rm max}^{\rm single}$ with $\ell=k$
(\ref{nmaxellk}) as a function of coalition size~$c$, at
$M=2^{24}$, $k=128$, $k_0=96$, $\qg=0.1$. 
The $L_{\rm suff}$ is set according to (\ref{codelength}), with $U/P_{\rm FP}=2^{30}$. 
}}
\label{fig:nmaxsingleell128}
\end{figure}

The parameters $M,k,k_0,\qg$, $L_{\rm suff}$ are usually given. 
That leaves $\ell$ to be optimised.
We observe (see Fig.\,\ref{fig:nmaxsingleell}) that it is advantageous to choose 
$\ell$ as large as possibe, i.e. $\ell =k$.
This can be understood intuitively:
Setting $w=1$ reduces the space occupied by a tracing entry to a single bit,
which is especially important when the coalition is large.
Eq.\,(\ref{nmaxsingleuse}) is slightly simplified to
\bea
	&& n_{\rm max}^{\rm single}(M,k,k_0,\ell=k,\qg,L_{\rm suff})
	\nn\\ && \quad\quad
	=\frac Mk-\frac{L_{\rm suff}}k
	\Big[1-{\rm InvBinoTail}(\ell,k_0,\qg)\Big]^{-2}.
\label{nmaxellk}
\eea
Fig.\,\ref{fig:nmaxsingleell128} shows an example of $n_{\rm max}$ (\ref{nmaxellk})
as a function of the coalition size~$c$.
We see that the reduction of $n_{\rm max}$ due to the increasing $L_{\rm suff}$
is rather mild.

\section{A scheme with re-use of blob entries}
\label{sec:withreuse}

\subsection{Scheme description}

\underline{Initialisation}\\
The Operator (pseudo)randomly generates $B_{[N]\setminus\cT}$. This data is the same for all users.
For each user $u$ individually he inserts tracing information $z^{(u)}$ in $B$ at locations~$\cT$.
Each user $u$ receives a personalised blob $B^{(u)}$. 

\underline{Encryption}\\
The Operator (pseudo)randomly generates a vector $L=(L_1,\ldots,L_\ell)\in([N]\setminus\cT)^\ell$.
(Note that for $i\neq j$, collisions $L_i=L_j$ are allowed.)
He encrypts the content using $B_L$ as key material.
He broadcasts the ciphertext and a compact description of~$L$ to all users.

\underline{Decryption}\\
User $u$ receives the ciphertext and the description of~$L$.
He decrypts the ciphertext using key material $B^{(u)}_L$.

\subsection{Statistics of the number of visited positions}

Let $n$ be the number of times that a key is derived from the blob.
Then $r=n\ell$ is the number of times that a random index is drawn.
Let $V\in([N]\setminus\cT)^r$ be the vector of drawn indices, $V=(V_1,\ldots,V_r)$.
Note that an index can occur in $V$ multiple times.
The attacker observes $V$.
We denote set of indices contained in $V$ as $\cV$, with $|\cV|\leq r$.

\begin{lemma}
\label{lemma:probV}
Let $s\leq r$. The probability distribution for the size $\cV$ is given by
\be
	\pr[|\cV|=s] 
	= \frac{(N-t)_s}{(N-t)^r}\stirling{r}{s}
	= \frac1{(N-t)^r}{N-t\choose s}s! \stirling{r}{s}.
\ee
\end{lemma}
\underline{\it Proof:}
The number of ways to choose $s$ out of $N-t$ positions, with ordering, is ${N-t\choose s}s!$;
the factor $\stirling{r}{s}$ is the number of partitions of $r$ into $s$ bins, such that no bin is left empty.
One has to divide by
the total number of ways to draw a vector of $r$ elements from $[N]\setminus\cT$, namely $(N-t)^r$. 
\hfill$\square$

\begin{lemma}
\label{lemma:visited}
The expected number of visited positions after $r$ steps is 
\be
	\EE_v |\cV|=(N-t)\Big[1-(1-\frac1{N-t})^r\Big].
\ee
\end{lemma}
\underline{\it Proof:}
From Lemma~\ref{lemma:probV} we have 
$\EE_v |\cV|=\sum_{s=0}^r\frac{(N-t)_s}{(N-t)^r}\stirling{r}{s} s$.
We write $\stirling{r}{s} s =\stirling{r+1}{s}-\stirling{r}{s-1}$. This yields two summations.
Both sums are evaluated using the sum rule
$\sum_{k=0}^n \stirling{n}{k}(x)_k = x^n$.

An alternative proof follows from counting non-visited positions.
Each position in $[N]\setminus\cT$ has probability $(1-\fr1{N-t})^r$ of not being visited. 
\hfill$\square$

\subsection{Analysis of the attack}

The attack is the same as in Section~\ref{sec:attacksingleuse}.
(Note that the attacker is now the only party that keeps track of~$\cV$.)

The relation between $t$ and $\qe_*$ is the same as in Section~\ref{sec:tracingsingleuse}.

The main difference w.r.t. the single-use scheme is that the 
failure probability of the attackers now depends on~$n$.

\subsubsection{Probability that the pirated blob fails to provide the next key}

The attackers randomly remove a fraction $\qe_*$ from $[N]\setminus\cV$. 

At fixed $\cV$,
the probability that a random blob index in $[N]\setminus\cT$ is not available in the pirated blob is 
given by
\be
	P_{\rm miss}\isdef \pr[\mbox{entry unavailable}|\cV]
	=\qe_*\frac{N-t-|\cV|}{N-t}.
\label{defPmiss}
\ee
Obviously, with increasing $|\cV|$ it becomes more likely that a random entry {\em is} available, since $\cV$ 
is fully kept in the pirated blob.
From (\ref{defPmiss}) we get a binomial tail expression for the probability that 
the number of unavailable entries is at least $k_0/w$, ensuring that brute forcing of the key is difficult,
\bea
	P_{\rm fail} 
	&\isdef &
	\pr[\#{\rm unavail.entries}\geq \frac{k_0}w] 
	\nn\\
	&=& \EE_{|\cV|} \sum_{a=\lceil k_0/w\rceil}^\ell {\ell\choose a}P_{\rm miss}^a(1-P_{\rm miss})^{\ell-a}
	\nn\\ &=&
	\EE_{|\cV|} {\rm BinoTail}(\ell,\lceil k_0/w\rceil,P_{\rm miss}).
\eea

\begin{lemma}
\label{lemma:failreuse}
Let $\qe_*\leq \frac1\ell\lceil \frac{k_0}w\rceil$.
Then
\be
	P_{\rm fail}\geq {\rm BinoTail}(\ell,\lceil k_0/w\rceil,\qe_*(1-\frac1{N-t})^{n\ell}).
\label{Pfailineq}
\ee
\end{lemma}
\underline{\it Proof:}
The expression ${\rm BinoTail}(\ell,\lceil k_0/w\rceil,P)$ viewed as a function of $P$
has a sigmoid shape, convex on $P\leq \frac1\ell\lceil k_0/w\rceil\approx k_0/k$ and concave for larger~$P$.
Via (\ref{defPmiss})
the condition $\qe_*\leq \frac1\ell\lceil \frac{k_0}w\rceil$ ensures that $P_{\rm miss}$ lies in the convex part.
We then use Jensen's inequality to write
$\EE_{|\cV|} {\rm BinoTail}(\ell,\lceil k_0/w\rceil,P_{\rm miss})\geq
{\rm BinoTail}(\ell,\lceil k_0/w\rceil,\EE_{|\cV|} P_{\rm miss})$
and finally invoke Lemma~\ref{lemma:visited}.
\hfill$\square$

\subsubsection{Figure of merit $n_{\rm max}$}

We will use the right hand side of (\ref{Pfailineq}) as an approximation for $P_{\rm fail}$.
For $\qe_*\leq \frac1\ell\lceil \frac{k_0}w\rceil$ we are then erring on the side of caution (Lemma~\ref{lemma:failreuse}).
However, also for larger $\qe_*$ the approximation is not bad, since the distribution of $|\cV|$
becomes narrow asymptotically.
The defining relation for the figure of merit $n_{\rm max}$ becomes
\be
	\qg = {\rm BinoTail}(\ell,\lceil\ell\frac {k_0}{k}\rceil,\qe_*[1-\fr1{N-t}]^{n_{\rm max}\ell}),
\label{threshold_reuse}
\ee
i.e. $n_{\rm max}$ is the value of $n$ where $P_{\rm fail}$ dives below the acceptable failure rate~$\qg$.
From (\ref{threshold_reuse}) we isolate $n_{\rm max}$,
\bea
	&& \hskip-4mm
	n_{\rm max}^{\rm multi}(M,k,k_0,\ell,\qg,L_{\rm suff},t) 
	\nn\\ &&
	=\frac{\ln[\qe_* / {\rm InvBinoTail}(\ell,\lceil\ell\frac {k_0}{k}\rceil,\qg)]}{-\ell \ln(1-\frac1{N-t})}
	\nn\\ && =
	\frac{\ln(1-\sqrt{L_{\rm suff}/t})-\ln {\rm InvBinoTail}(\ell,\lceil\ell\frac {k_0}{k}\rceil,\qg)}
	{-\ell \ln(1-\frac1{N-t})}.
	\quad\quad
\label{nmaxmultifull}
\eea
(Here $N=M/w$.)
Expanding the logarithm in the denominator up to first order we get the approximation
\bea
	&& \hskip-10mm
	n_{\rm max}^{\rm multi}(M,k,k_0,\ell,\qg,L_{\rm suff},t) 
	\nn\\ && \approx
	\frac{N-t}{\ell}
	\ln\frac{1-\sqrt{L_{\rm suff}/t}}{{\rm InvBinoTail}(\ell,\lceil\ell\frac {k_0}{k}\rceil,\qg)}
	\nn\\ &&=
	(\frac M k-\frac t\ell)
	\ln\frac{1-\sqrt{L_{\rm suff}/t}}{{\rm InvBinoTail}(\ell,\lceil\ell\frac {k_0}{k}\rceil,\qg)}
\label{nmaxreuse}
\eea
with relative error of order $\fr1{N-t}$.
Comparing (\ref{nmaxreuse}) to (\ref{nmaxsingleuse}), we see that the re-use of blob entries gives rise to a
logarithmic factor.
Furthermore, the parameter $t$ can be freely chosen in the re-use case, while it is fixed in the single-use case.

\subsubsection{Setting the parameters}

\begin{figure}[t]
\begin{center}
\includegraphics[width=70mm]{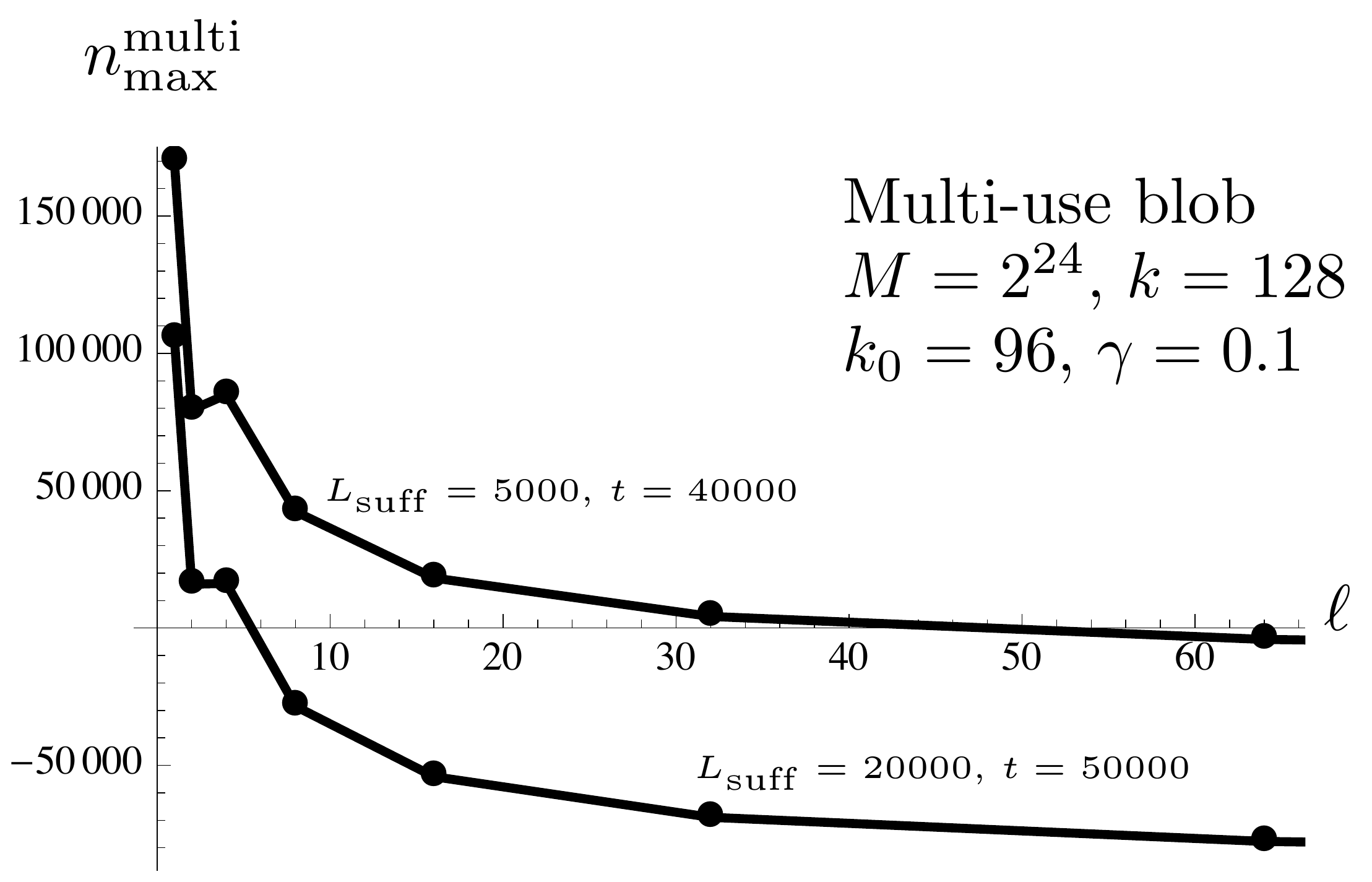}
\end{center}
\caption{ {\it
$n_{\rm max}^{\rm multi}$ (\ref{nmaxmultifull}) as a function of $\ell$, at
$M=2^{24}$, $k=128$, $k_0=96$, $\qg=0.1$. 
Top curve: $L_{\rm suff}=5000, t=40000$. Bottom curve: $L_{\rm suff}=20000,t=50000$. 
}}
\label{fig:nmaxmultiell}
\end{figure}

Again we consider the parameters $M,k,k_0,\qg,L_{\rm suff}$ to be fixed.
The task is to maximise $n_{\rm max}^{\rm multi}$ as a function of 
the still free parameters $\ell$ and~$t$.
We observe (see Fig.\,\ref{fig:nmaxmultiell}) that the best results are obtained with $\ell=1$.
Note that large values of $\ell$ can cause (\ref{nmaxmultifull}) to produce negative\footnote{ 
Erring on the side of safety too much in Lemma~\ref{lemma:failreuse}.
}
$n_{\rm max}$, 
especially when the coalition is large.
The intuitive explanation is that at large $\ell$ the colluders quickly learn many functional indices;
in contrast to the single-use case, this knowledge is helpful for them
since it allows them to build a pirated blob with many functional entries.
In the special case $\ell=1$ the expression ${\rm BinoTail}(\ell,\lceil \ell k_0/k\rceil,P_{\rm miss})$
reduces to $P_{\rm miss}$, and the result for $n_{\rm max}$ simplifies to
\bea
	&& \hskip-12mm
	n_{\rm max}^{\rm multi}(M,k,k_0,\ell=1,\qg,L_{\rm suff},t) 
	\nn\\ &&
	=\frac{\ln(1-\sqrt{L_{\rm suff}/t})+\ln \frac1\qg}{- \ln(1-\frac1{M/k-t})}
\label{multianalytic}
	\\ &&=
	\Big[\frac Mk-t-\cO(1)\Big]\ln\frac{1-\sqrt{L_{\rm suff}/t}}\qg.
\label{nmaxell1}
\eea
Note that we have the constraint $t>L_{\rm suff}/(1-\qg)^2$.

\begin{figure}[h]
\begin{center}
\includegraphics[width=80mm]{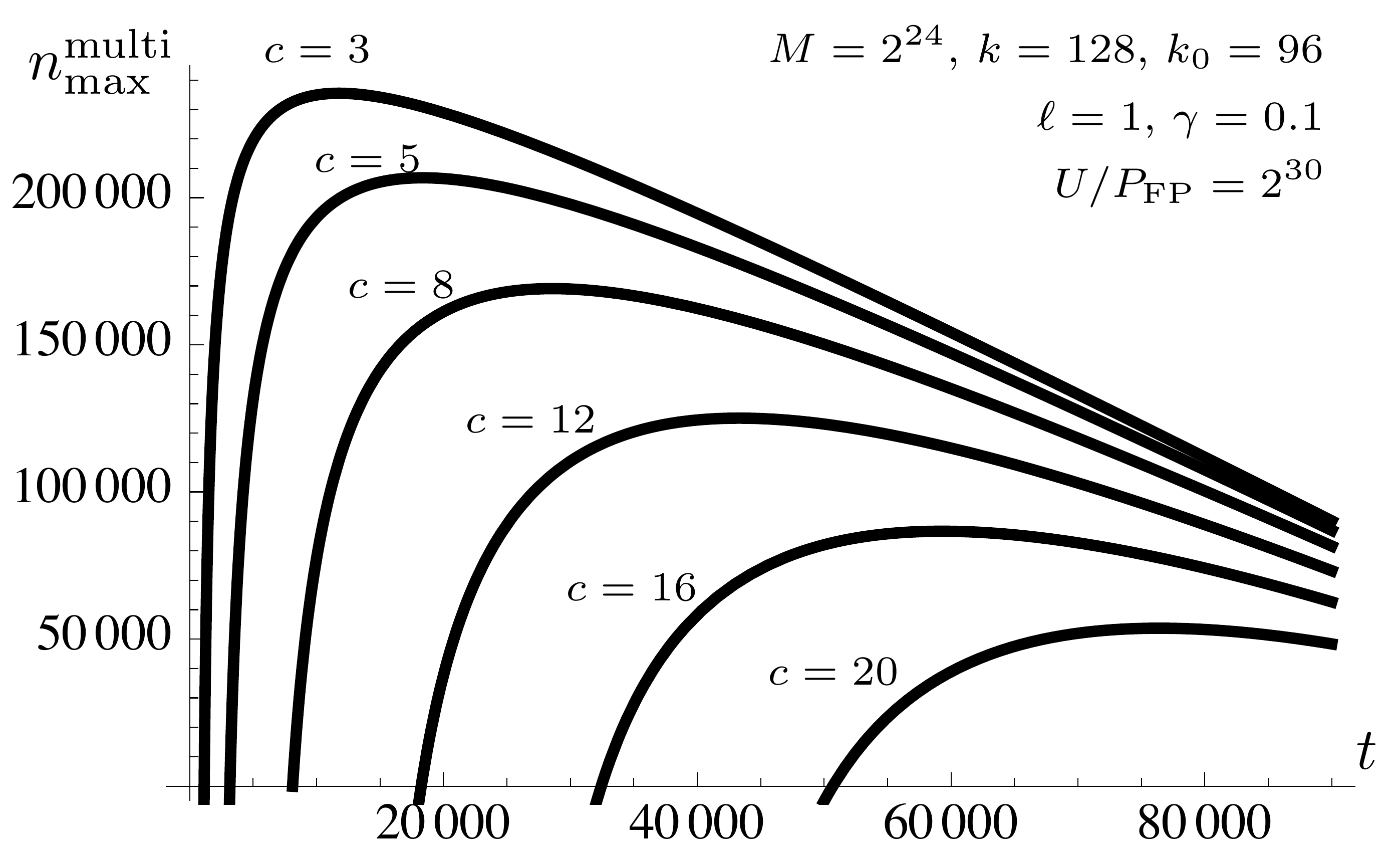}
\end{center}
\caption{ {\it
$n_{\rm max}^{\rm multi}$ (\ref{multianalytic}) as a function of $t$, at
$M=2^{24}$, $k=128$, $k_0=96$, $\ell=1$, $\qg=0.1$ for various coalition sizes~$c$.
The $L_{\rm suff}$ is set according to (\ref{codelength}), with $U/P_{\rm FP}=2^{30}$. 
. 
}}
\label{fig:nmaxmultit}
\end{figure}

\section{Optimisation for the pay-TV use case}
\label{sec:payTV}

A consumer in a pay-TV system uses a decoder to access content. 
To prevent unauthorized access to the content, it is encrypted at the head-end system 
of the pay-TV provider before it is distributed to the decoders, and only authorized 
decoders are given access to the content decryption keys. 
A pay-TV system adopts a broadcast model for distributing the encrypted content and 
the associated key management messages, sending the same encrypted content and messages 
to all decoders. 
There is usually a limited amount of bandwidth available for distributing key management 
messages since the provider prefers to use as much as possible of the available 
bandwidth for distributing content. 
For this reason, a pay-TV security architecture typically contains keys that are shared 
between a number of decoders. 
For detailed information and a concrete example we refer to \cite{KMR07,R11,WGL1998}.

If a shared key is compromised from a decoder and re-distributed, 
then the provider wants to trace the decoder or decoders that have been compromised 
and revoke these decoders to halt the piracy. 
Our scheme may be applied to watermark a high-level node key (e.g.~the root key) in a revocation tree as used in \cite{R11,WGL1998}, 
so that publishing this key becomes risky for pirates, even if they do a collusion attack on the watermark. 
The corresponding blob can be pre-distributed to each decoder during manufacturing.

Below we investigate the relation between the size of a 
blob and the maximum number of colluders that can be traced successfully. 
This number relates directly to the business case of the adversary. 
For example, if compromising the blob from the read-proof and tamper-resistant 
security chip in the decoder costs \$50,000 per chip (i.e.\,per decoder) 
and if the scheme is resistant against 8 colluders, then the costs of a
corresponding attack that cannot be corrected by the provider ``over-the-air'' is at least \$450,000.

For a concrete example, we use the values in Table~\ref{table:pay-TV}. 
The number of blob uses assumes that the lifetime of a decoder is 7 years and that 
the key is updated every 30 minutes. 
Note that a large value of $P_{\rm FP}$ is allowed: 
the revocation of a decoder based on a false positive can 
easily be reversed if necessary, e.g. after receiving a service call from the corresponding customer.

\begin{figure}[!h]
\begin{center}
\includegraphics[width=70mm]{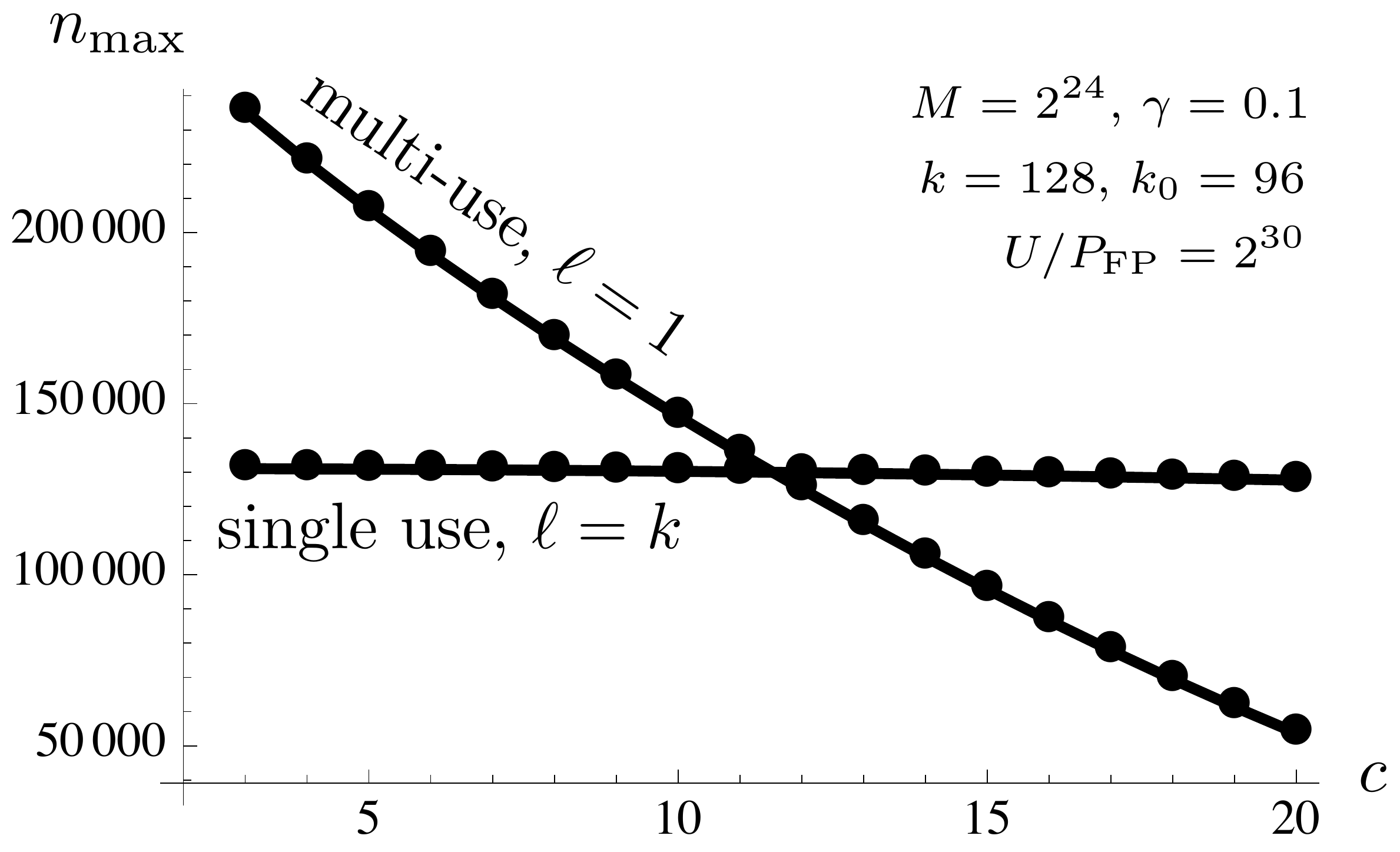}
\end{center}
\vskip-5mm
\caption{ {\it
Comparison of $n_{\rm max}$ in the single-use and multi-use case.
In the multi-use case the parameter $t$ is set to its optimal value.
$M=2^{24}$, $k=128$, $k_0=96$, $\qg=0.1$. 
The $L_{\rm suff}$ is set according to (\ref{codelength}), with $U/P_{\rm FP}=2^{30}$. 
}}
\label{fig:combo}
\end{figure}

In Fig.\,\ref{fig:combo} we compare the figure of merit $n_{\rm max}$
for the single-use and multi-use case using the numbers from Table~\ref{table:pay-TV}.
\begin{itemize}[leftmargin=4mm,itemsep=0mm]
\item
We observe a crossover.
For small coalitions the multi-use scheme performs better than the single-use scheme,
whereas this is reversed for large coalitions.
The crossover is understood as follows.
On the one hand, re-using blob entries has the advantage that it allows, in principle,
for a large~$n$.
On the other hand, re-use has the drawback that the attackers learn entries that 
they can incorporate into the pirated blob without fear of being traced.
With increasing coalition size the drawback becomes more pronounced,
since the ratio of functional entries versus tracing entries in the pirated blob
increases.
As we saw in Section~\ref{analysissingle}, the $n_{\rm max}$ for the single-use scheme
depends on $c$ relatively weakly.
\item
The achievable $n_{\rm max}$ is in line with the desired number of uses in the pay-TV context
($\approx 1.2\cdot10^5$).
\end{itemize}

\renewcommand{\arraystretch}{1.2}
{\small
\begin{table}[h]
\caption{\it Typical values for the parameters in pay-TV systems.}
\begin{center}
\begin{tabular}{ | l | l | c | }
\hline
$M$			& Size of the blob, in bits.						& preferably $\leq 2^{24}$	\\ \hline
$k$			& Key size, in bits.							& 128 \\ \hline
$k_0$		& Key size difficult to brute force. 	& 96 \\ \hline
$n$ 			& Number of blob uses.						& preferably 
\\ &&
$\geq 7 \cdot 365 \cdot 24 \cdot 2 \approx 2^{17}$ \\ \hline
$U$ 			& Number of users.							& $2^{20}$ \\ \hline
$c_0$ 			& Anticipated number of attackers.						& preferably $\geq 8$ \\ \hline
$P_{\rm FP}$	& False Positive probability.					& $ \leq 2^{-10}$ \\ \hline
$L_{\rm suff}$	& Length of binary Tardos code.					& $\geq 6.6\cdot10^3$ \\ \hline
\end{tabular}
\end{center}
\label{table:pay-TV}
\end{table}
}

\section{Discussion}
\label{sec:discussion}

We have introduced and analysed two BKC schemes for symmetric decryption and traceability in the whitebox attacker model,
focusing on pay-TV scenarios that involve keys shared by multiple customers.
Compared to WBC, this approach has the advantage that incompressibility and traceability are
achieved in an information-theoretically secure way.
It may also be interesting to use our schemes instead of group signatures,
as a difficult-to-extricate group credential with post-quantum security.

We have shown that blob schemes are feasible in the context of pay-TV. 
The parameter values are consistent with typical pay-TV requirements.
For the single-use scheme with large $\ell$, special care must be taken to keep the control
messages small.
A blob scheme may be used as a way to watermark node keys close to the root,
so that publishing such keys becomes risky for pirates, even if they do a collusion attack on the watermark.

\vskip2mm

Our schemes can be tweaked and adapted in various ways.
There is a large space from which the design parameters 
$M,k,k_0,\qg,c_0$ can be chosen.
We also mention that the parameter $t$ (the number of tracing positions) can be kept secret.
This would force the colluders to make worst-case assumptions and hence discard a larger fraction $\qe$
than if they had known~$t$. 

Furthermore, it is possible to use {\em asymmetric} cryptography.
Consider a set of position-dependent moduli $N_i=p_i q_i$, where the factorisation is known only to the Operator.
The Operator broadcasts $(i, N_i, a)$, where $i$ is a pointer to blob entry $b_i$ and $a$ is a number.
The user computes the decryption key as $a^{b_i}{\rm mod}\,N_i$.
Watermarking is achieved by handing out to different users
blob entries $b_i$ that differ by an integer multiple of $\qf(N_i)=(p_i-1)(q_i-1)$.
In this way it becomes possible to watermark {\em functional} blob entries, i.e.~{\em all} blob entries. 
We leave such schemes as a topic for future work.


\section*{Acknowledgment}
Part of this research was funded by NWO grant 14648.


\bibliographystyle{plain}
\bibliography{incompressible}

\end{document}